\documentclass[twocolumn,pra,showpacs,superscriptaddress]{revtex4}

\usepackage{graphicx}
\usepackage{dcolumn}
\usepackage{bm}
\usepackage{amssymb}
\usepackage{amsmath}
\usepackage{extarrows}

\begin{document}


\title{Simulation and measurement of the fractional particle number in one-dimensional optical lattices}

\author{Dan-Wei Zhang}
\email{zdanwei@hku.hk}
\affiliation{Department of Physics and Center of Theoretical and Computational Physics, The University of Hong Kong,
Pokfulam Road, Hong Kong, China}

\author{Feng Mei}
\affiliation{National Laboratory of Solid
State Microstructures and School of Physics, Nanjing University,
Nanjing 210093, China}

\author{Zheng-Yuan Xue}
\affiliation{Guangdong Provincial Key  Laboratory
of Quantum Engineering and Quantum Materials, SPTE, South China
Normal University, Guangzhou 510006, China}\affiliation{Department of Physics and Center of Theoretical and
Computational Physics, The University of Hong Kong, Pokfulam Road,
Hong Kong, China}

\author{Shi-Liang Zhu}
\email{slzhu@nju.edu.cn} \affiliation{National Laboratory of Solid
State Microstructures and School of Physics, Nanjing University,
Nanjing 210093, China}

 \affiliation{Synergetic Innovation Center of Quantum Information and Quantum
Physics, University of Science and Technology of China, Hefei
230026, China}

\author{Z. D. Wang}
\email{zwang@hku.hk} \affiliation{Department of Physics and Center of Theoretical and Computational Physics, The University of Hong Kong,
Pokfulam Road, Hong Kong, China}

\begin{abstract}
We propose a scheme to mimic and directly measure the
fractional particle number in a generalized Su-Schrieffer-Heeger
model with ultracold fermions in one-dimensional optical lattices.
We show that the fractional particle number in this model can be
simulated in the momentum-time parameter space in terms of Berry
curvature without a spatial domain wall. In this simulation, a
hopping modulation is adiabatically tuned to form a kink-type
configuration and the induced current plays the role of an
analogous soliton distributing in the time domain, such that the
mimicked fractional particle number is expressed by the particle
transport. Two feasible experimental setups of optical lattices
for realizing the required Su-Schrieffer-Heeger Hamiltonian with
tunable parameters and time-varying hopping modulation are
presented. We also show practical methods for measuring the
particle transport in the proposed cold atom systems by
numerically calculating the shift of the Wannier center and the
center of mass of an atomic cloud.
\end{abstract}

\date{\today}

\pacs{03.75.Ss, 03.67.Ac, 03.65.Vf}

\maketitle

\section{introduction}

Particle fractionalization has been recognized as a remarkable and
fundamental phenomenon in both relativistic quantum field theory
and condensed matter systems
\cite{JR,Niemi,SSH,SSHreview,Laughin,Qi2008a,Xiao2010,2D-FPN,Exp-FQHE,Exp-1DIE}. The first
physical demonstration of fractionalization is the celebrated
Su-Schrieffer-Heeger (SSH) model of one-dimensional (1D) dimerized
polymers \cite{SSH,SSHreview}, such as polyacetylene. In this
model, a kink domain wall in the electron hopping configuration
induces a zero-energy soliton state carrying a half-charge
\cite{SSH,SSHreview}. The basic physics of fractionalization in
SSH model is governed by a low-energy effective Dirac
Hamiltonian with topologically nontrivial background fields,
which was firstly proposed by Jackiw and Rebbi \cite{JR,Niemi}.
Subsequent achievements were made to generalize the original
SSH model to exhibit an irrational (arbitrary) fermion number by
breaking the conjugation symmetry
\cite{Goldstone,Rice,Jackiw1983}. The fractional particle number
(FPN) in these systems can be understood in terms of global
deformations of the hole sea (or the valence band) due to the
nontrivial background fields.

SSH model has achieved great success in describing
transport properties of polymers, and some novel phenomena
associated with the topological solitons have also been explored
in experiments \cite{SSHreview}. However, the intrinsic FPN
has never been experimentally detected due to the spin-doubling
problem in these materials: two spin orientations are present for
each electron and thus a domain wall in the polyacetylene carries an
integer charge \cite{SSHreview}. Inspired by the newly discovered
quantum spin Hall insulators \cite{TI}, it was theoretically
proposed to realize SSH model in an edge of this
two-dimensional insulator by bringing a magnetic domain wall there,
and the edge electrons with the inherent chiral symmetry may
provide a direct signature of FPN \cite{Qi2008b,Ojanen}. However,
creation of such a magnetic domain wall acting only on the edge
elections is experimentally challenging and the proposed schemes
are yet to be demonstrated.

In the past years, a lot of theoretical and experimental work has
been carried out to simulate the Dirac equation and the involved
exotic effects by using ultracold atoms
\cite{Zhu2007,Goldman,Vaishnav,Zhu2009,DEreview,KTexp,Tarruell,ZBexp1,ZBexp2}.
Especially, it has been proposed to realize the (generalized) SSH
model associated with effective Dirac Hamiltonian using ultracold
atomic gases in the continuum \cite{zhang2012} and in optical
lattices \cite{Ruostekoski1,Ruostekoski2,liu2013,li2013}. The
detection of FPN in these cold atom systems was also suggested by
optical image of the density distribution of soliton modes
\cite{zhang2012,Ruostekoski2}. Since single-component fermionic gases or component-dependent optical lattices are used in the realization of FPN in atomic systems, the spin-doubling problem encountered in condensed matter systems can be avoided. In a recent experiment with a 1D
optical superlattice, SSH model in the absence of spatial domain walls
was realized and its topological features were also probed
\cite{Bloch2013}, making direct measurement of FPN in optical
lattices to be feasible and timely.

In this work, we propose a new scheme to mimic and directly
measure FPN in the generalized SSH model  using ultracold
fermions in 1D optical lattices. Firstly, we show that FPN in
this model can be simulated in the momentum-time parameter space
in terms of Berry curvature without creating a spatial domain
wall. In this simulation, a hopping modulation is adiabatically
tuned to form a kink-type configuration and the induced current
plays the role of an analogous soliton distributing in the time domain, so that
the mimicked FPN in parameter space is expressed by the adiabatic
particle transport. Furthermore, we explore how to implement
this new scheme with ultracold fermions in 1D optical lattices. We
propose two experimentally setups to realize the required SSH
Hamiltonian with tunable parameters and hopping modulations, and
then show practical methods for measuring the particle transport
in the proposed cold atom systems by numerically calculating the
shift of the Wannier center and the center of mass of an atomic
cloud. Some possible concerns in realistic experiments, such as
the energy scales, the adiabatic condition and the effects of an external harmonic trap, are also
considered. In comparison with the previous proposals of realizing
and detecting FPN
\cite{Qi2008b,zhang2012,Ruostekoski1,Ruostekoski2,liu2013,li2013},
one advantage of the presented scheme is that it does not require
the spatial domain in the hopping configuration, which is usually
hard to create. Another advantage is the adiabatic particle
transport corresponding to the value of PFN can be directly
measured in our proposed cold atom systems.

The rest of this paper is organized as follows: Section II presents a brief
review on fractionlization in a generalized SSH model. In Sec.
III, we elaborate our scheme of simulation and measurement of
FPN in this model, based on the Berry curvature and adiabatic
transport approaches. In section IV, we propose two feasible
experimental setups of 1D optical lattices to realize the required
Hamiltonian, and then discuss how to measure the atomic particle
transport in the proposed systems. Finally, a short conclusion is
given in Sec. V.

\section{Fractionalization in SSH model}

Before describing our scheme, we briefly review the arbitrary FPN in the generalized SSH model in this section. We start with this model 
described by a tight-binding Hamiltonian \cite{Rice,Xiao2010}
\begin{eqnarray}
\nonumber &&H=\sum_{n}\left[ J+(-1)^n\delta\right]\left(\hat{c}_n^{\dag}\hat{c}_{n+1}+\text{H.c.}\right)
\\&&~~~~~~~+\frac{\Delta}{2}\sum_n(-1)^n \hat{c}_n^{\dag}\hat{c}_{n},
\end{eqnarray}
where $\hat{c}_n$ ($\hat{c}_n^{\dag}$) is the fermion annihilation (creation) operator in site $n$, $J$ is the uniform hopping amplitude, $\delta$ is the
dimerized hopping modulation, and $\Delta$ is a staggered potential breaking the inversion symmetry [the conjugation symmetry in the low-energy Dirac Hamiltonian (3)]. In this lattice system, as shown in Fig. 1(a), the even and odd number sites form two sublattices with modulating hopping amplitudes (on-site energies) and thus a unit cell contains two nearest lattice sites, which are used to constitute a pseudo-spin. When the inversion (conjugation) symmetry preserves with $\Delta=0$, the system corresponds to the original SSH model of polyacetylene.

\begin{figure}[tbp]
\includegraphics[width=7cm]{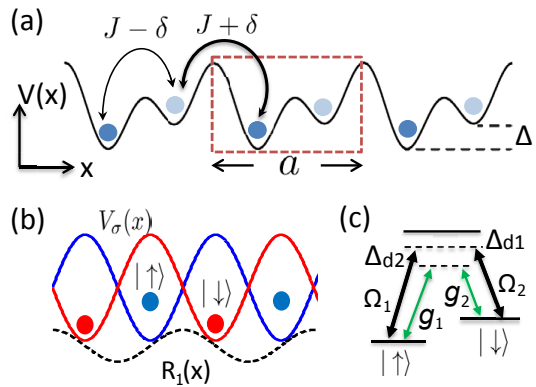}
\caption{(Color online) Two experimental setup for simulating SSH Hamiltonian in 1D optical lattices. (a) A double-well optical superlattice trapping noninteracting single-component fermionic atoms. A unit cell contains two nearest lattice sites with energy offset $\Delta$ (for atoms denoted by dark and light blue small balls) and the atomic hopping exhibits staggered modulation configuration. (b) A state-dependent optical lattice trapping noninteracting two-component fermionic atoms, where atomic states $|\uparrow\rangle$ and $|\downarrow\rangle$ are denoted by blue and red small balls. The block dotted line represents a Raman field $R_1(x)$. (c) Raman-assisted tunneling. The uniform atomic hopping between the nearest-neighbour in the state-dependent optical lattice is realized by a large-detuned Raman transition with detuning $\Delta_{d1}$ and Rabi frequencies of the Raman beams $\Omega_{1,2}$, while the hopping modulation is realized by another pair of Raman beams $g_{1,2}$ with large detuning $\Delta_{d2}$. The Zeeman splitting gives rise to the $\Delta$ term in this system.}
\end{figure}

By employing Fourier transformation in the spin basis, we can obtain the Bloch Hamiltonian of the model as
\begin{equation}
\mathcal{H} = \vec{d}(k)\cdot\vec{\sigma},
\end{equation}
where $\vec{\sigma}=(\sigma_x,\sigma_y,\sigma_z)$ are the Pauli matrices acting on the pseudo-spin, $\vec{d}(k)=(J\cos\frac{ka}{2},-\delta\sin\frac{ka}{2},\Delta)$ is the band vector with $a$ being the lattice spacing as shown in Fig. 1(a). By linearizing the Bloch bands near the Dirac point $k_D=\pi/a$, Hamiltonian (2) can be transformed into an effective low-energy relativistic Hamiltonian \cite{JR,Niemi,Goldstone}
\begin{equation}
H_D=v_F\hat{p}_x\sigma_x-2\delta\sigma_y+\frac{\Delta}{2}\sigma_z
\end{equation}
in the continuum, where $v_F=Ja/\hbar$ is the Fermi velocity, $\hat{p}_x$ is the momentum operator measured from the Dirac point, $\delta$ and $\Delta$ act as two background fields \cite{Goldstone}.

It has been widely studied that for a kink-type background potential with $\delta(x\rightarrow\pm\infty)=\pm\delta_0$, an unpaired soliton state appears at the kink carrying FPN \cite{Niemi,Goldstone}
\begin{equation}
\mathcal{N}_s=-\frac{1}{\pi}\arctan(\frac{4\delta_0}{\Delta}),
\end{equation}
which may exhibit arbitrary fractional eigenvalues. The minus sign in FPN is due to the fact that the physical
fermion number in the soliton sector is defined as being measured relative to the free sector without the kink background, and this fractional part of the fermion number actually comes from the global contribution (polarization) of the valence band \cite{Niemi,Goldstone,zhang2012}. In addition, it has a topological character in the sense that it is dependent only on the asymptotic behavior of the background fields instead of their local profiles. When $\Delta\rightarrow0$, it recovers to the half fermion number $\pm\frac{1}{2}$ for the zero-energy soliton mode in the original SSH model \cite{JR}.

It is interesting to note that fractionalization also exhibits in many low-dimensional correlated electron systems. For instance, a well-known example is that of fractional excitations in the fractional quantum Hall regime \cite{Laughin,Exp-FQHE}, which is a consequence of strong Coulomb interaction among 2D electrons in partially filled Landau levels. In addition, the collective excitations in some 1D interacting fermion systems amay be characterized by effective fractional charges via the spin-charge separation mechanism \cite{Exp-1DIE}. Fractionalization in these systems is basically due to election correlations and hence is completely different from that in SSH model of noninteracting fermions. The fractional charges in some correlated electron systems have been directly observed in experiments \cite{Exp-FQHE,Exp-1DIE}. However, despite the fractionalization in SSH model being investigated for decades, FPN there is yet to be directly measured in solid state materials (mostly due to the spin-doubling problem) or in artificial systems, even for the simplest half-fermion-number case with $\Delta=0$. Therefore, experimentally feasible schemes for direct measurement of the intrinsic FPN in SSH model would be of great value.

\section{scheme to directly measure PFN}

Several schemes have been proposed to realize SSH model and to
probe the soliton modes with FPN in cold atom  systems
\cite{Ruostekoski1,Ruostekoski2,liu2013,li2013}; however, the
realization of the required kink background and the local
detection of the fermion number of a soliton state therein are still
challenging in practical experiments. In the following, we
propose a simpler scheme to mimic and then directly measure PFN
via the adiabatic particle transport.

The energy bands of SSH model described by Hamiltonian (2) can
be mapped onto a two-level system in the Bloch sphere with the
parameterized Bloch vector
$\vec{S}=h_0(\sin\theta\cos\phi,\sin\theta\sin\phi,\cos\theta)$,
where $\theta$ and $\phi$ are respectively the polar and azimuthal
angles. In this mapping, we have
\begin{equation}
\begin{array}{ll}
h_0 = \sqrt{4J^2\cos^2(ka/2)+4\delta^2\sin^2(ka/2)+\Delta^2/4},\\
\theta = \arccos(\Delta/2h_0),\\
\phi = -\arctan\left[\frac{\delta\sin(ka/2)}{J\cos(ka/2)}\right].
\end{array}
\end{equation}
The degeneracy point locates at $k=\pi/a$, $\delta=0$ and
$\Delta=0$. We consider the fractionalization in this band
insulator  at half filling \cite{zhang2012}, corresponding to the
low-energy level with the eigenstate
$|u_{-}\rangle=(\sin\frac{\theta}{2}e^{-i\phi},-\cos\frac{\theta}{2})^{\text{T}}$
with T being the transposition of matrix. In this framework, one
can define the Berry connection $A_{\theta}=\langle
u_{-}|i\partial_{\theta}u_{-}\rangle=0$ and $A_{\phi}=\langle
u_{-}|i\partial_{\phi}u_{-}\rangle=\sin^2\frac{\theta}{2}$.

\begin{figure}[tbp]
\includegraphics[width=8cm]{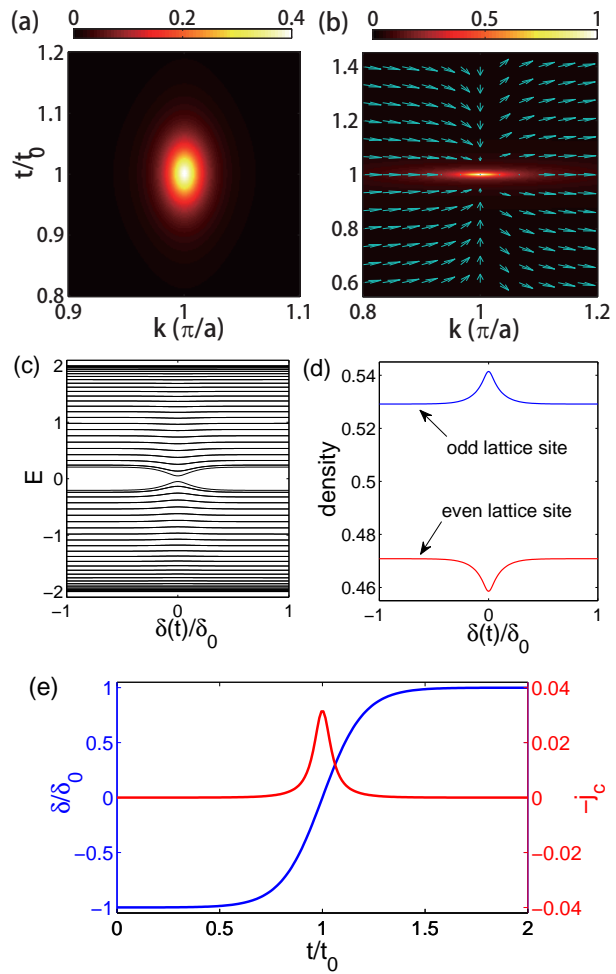}
\caption{(Color online) (a) Berry curvature distribution of the ground band $\mathcal{F}_{kt}$ in the center region of the $k$-$t$ parameter space. The Berry curvature outside is almost vanishing, and integration over the whole parameter space $(0\leqslant k\leqslant 2\pi/a,0\leqslant t\leqslant 2t_0)$ gives the particle transport. (b) Spin texture of SSH model, interpreted as a mapping from the $k$-$t$ parameter space onto the Bloch sphere by Eq. (5). The colors show the $S_z$-component $\cos[\theta(k,t)]$ and the arrows show the azimuthal component $(S_x,S_y)=(\sin\theta\cos\phi,\sin\theta\sin\phi)$.
Here the lengths of all the arrows has been divided by a factor $\pi$ for visibility.
(c) The energy spectrum for a lattice system with size $L=100$. (d) The variation of density in each lattice site with respect to the time-varying hopping modulation at hall filling. (e) The analog of kink potential and soliton-like current in the time domain under adiabatic conditions. The parameters in (a-e) are $J=1$, $\delta_0=\Delta=0.1$ and $t_0=5/\xi$.}
\end{figure}

Instead of considering the kink background field in the spatial
domain, here we introduce a time-varying hopping modulation with
a kink-type ramping configuration
\begin{equation}
\delta(t)=\delta_0\tanh\left[\xi(t-t_0)\right],
\end{equation}
where $t_0$ denotes the center of the time domain wall and $\xi$
represents the ramp frequency.  We assume $t_0\gg1/\xi$ (actually
$t_0=5/\xi$ is large enough) such that $\delta(0)\simeq -\delta_0$
at the beginning $t=0$, and then adiabatically ramp the system to
$t=t_f=2t_0$ with $\delta(t_f)\simeq \delta_0$ at the end. That is
to say, we simulate a kink potential in the time domain instead of
creating it in real space. In this dynamical case, the bulk gap is
$E_g=2\sqrt{\Delta^2+16\delta(t)^2}$, which will close at $t=t_0$
for original SSH model with $\Delta=0$. To guarantee the adiabatic
condition for original SSH case, which requires a gapped
bulk band, we can use a time-varying staggered potential with the
form $\Delta(t)=\delta_0\sin(\pi t/2t_0)$.

We consider the adiabatic evolution of the system  with the
ramping parameter $\delta(t)$ (and $\Delta(t)$ for original SSH
case) and provide that the Fermi level lies inside the band gap in
the whole progression. The Berry phase effect of this 1D band insulator can
be measured from the particle transport \cite{Xiao2010}. This is
an analog of the adiabatic charge pumping proposed by Thouless
\cite{Thouless}; however, the parametric driving in our case does
not form a closed cycle but only a half one. The topological
pumping in cold atom systems and photonic quasi-crystals has been
discussed in the contexts of SSH model \cite{TP1,TPGuo,TP2} and 1D
quasi-periodic Harper model \cite{Harper,TPphoton,TP3,TP4,TP5},
where the pumping particle is shown to be quantized one over one
period and can be
fractional over a fraction of one period \cite{TP5}.

In the momentum-time ($k$-$t$) parameter space, we can rewrite the
Berry connection  as $A_{k}=\partial_k\phi
A_{\phi}+\partial_k\theta A_{\theta}$ and $A_{t}=\partial_t\phi
A_{\phi}+\partial_t\theta A_{\theta}$. Thus the Berry curvature
$\mathcal{F}_{kt}=\partial_k A_t-\partial_t A_k$ in the $k$-$t$
space can be obtained as
\begin{equation}
\mathcal{F}_{kt}=\frac{\Delta J \sin^2\frac{ka}{2}\partial_t\delta(t)}{2[4J^2\cos^2\frac{ka}{2}+4\delta(t)^2\sin^2\frac{ka}{2}+\Delta^2/4]^{\frac{3}{2}}},
\end{equation}
where $J$ and $\Delta$ are assumed to be constants here. We note
that Berry curvature distribution given by Eq. (7) is modified for
time-varying $\Delta(t)=\delta_0\sin(\pi t/2t_0)$ discussed
previously, however, the corresponding PFN given by following
equation (8) remains since it just depends on the boundaries of
the background fields \cite{SSHreview}.

Figure 2(a) shows an example of the Berry curvature distribution
in the center region of the $k$-$t$ space for typical parameters,
while the Berry curvature outside is almost vanishing. In small
$\Delta$ limit, a sharp peak  exhibits in the Berry curvature
distribution at the position of $(k,t)=(\pi/a,t_0)$, which is the
dominant contribution to its integration over the parameter space.
Figure 2(b) shows the corresponding spin texture, which is
interpreted as a mapping from the $k$-$t$ parameter space onto the
Bloch sphere by Eq. (5). Figures 2(c) and 2(d) show the static
energy spectrum and the adiabatic density variation in each
lattice site with respect to the hopping modulation for the lattice
size $L=100$, respectively. It looks like a density kink-soliton
(anti-soliton) configuration appears in the time-domain.

The particle transport for the ground band in this 1D band insulator over the ramping progression of parameter $\delta(t)$ from $t=0$ to $t=t_f$ is given by the integration of the Berry curvature
\begin{eqnarray}
\nonumber && Q = -\frac{1}{2\pi}\int_0^{t_f} dt \int_{0}^{2\pi}dk\mathcal{F}_{kt}\\&&~~=\frac{1}{2\pi}\int_0^{2\pi}dk \left[A_k(k,t_f)-A_k(k,0)\right]\\ \nonumber &&~~=-\frac{1}{\pi}\arctan\left(\frac{4\pi J\delta_0}{\Delta\sqrt{\pi^2J^2+4\delta_0^2+\Delta^2/4}}\right)\\ \nonumber &&~~\simeq -\frac{1}{\pi}\arctan\left(\frac{4\delta_0}{\Delta}\right)=\mathcal{N}_s,
\end{eqnarray}
where the approximation satisfies well for $J\gg\delta_0,\Delta$, and becomes exact when $\Delta=0$. In the calculation, the integration of $\partial_k A_t$ over $k$ vanishes due to the periodic condition in the Berry vector potential \cite{Xiao2010}. Here the unquantized adiabatic particle can be regarded as the polarization change in this 1D band insulator \cite{Vanderbilt,Xiao2010}, which has also been discussed in the context of nanotubes and ferroelectrics materials \cite{Mele}. If one consider an additional anti-kink-type modulation to form a full cycle, then the particle transport will be quantized ($\pm1$ or $0$ depending on the loop of the cycle) after one period as the integration of the Berry curvature in the extensional region contribute the other fractional portion \cite{Qi2008b,Xiao2010}.

There are actually close connections between FPN of a soliton state and the adiabatic transport via the Berry phase approach in Eq. (8). We can consider the response equation in the progression of dynamical generation of the background field \cite{Goldstone,Qi2008a}:
\begin{eqnarray}
\rho_s=\frac{1}{2\pi}\partial_x\Theta(x,t),~~j_c=\frac{1}{2\pi}\partial_t\Theta(x,t),
\end{eqnarray}
where $\Theta(x,t)$ represents the generalized angular angle of the background field, $\rho_s$ denotes the soliton density distribution near the spatial domain wall, and $j_c$ is the induced current. In the present model with $J\gg\delta,\Delta$ (in which case the same polarization variation is obtained by the band Hamiltonian and the low-energy Dirac Hamiltonian), the angular angle $\Theta=-\arctan(4\delta/\Delta)$ is just time-dependent. Therefore, the induced current mimics an analog of kink-soliton in the time domain, as shown in Fig. 2(e). The induced current over the whole time domain gives the transferred particle
\begin{eqnarray}
Q=\int_0^{t_f} j_c(t) dt,
\end{eqnarray}
which takes the value given by Eq. (8) and depends only on the boundaries of the angular angle under the adiabatic condition.

So far, we have described our scheme to simulate FPN in SSH model in a parameter space and to directly measure it via the adiabatic transport. In contrast to the previous schemes \cite{Qi2008b,zhang2012,Ruostekoski1,Ruostekoski2,liu2013,li2013}, the presented scheme does not involve the spatial kink domain in the hopping configuration, which is usually hard to realize and (or) control in experiments. In addition, the particle transport corresponding to the value of PFN can be directly measured in cold atom systems, such as from the measurement of atomic density distribution and atomic current \cite{Bloch2014}, which will be discussed in 1D optical lattices in the next section.

\section{experimental implementation in optical lattices}

In this section,  we turn to discuss the implementation of our
scheme of mimicking and measuring FPN in 1D optical lattices.
We first propose two experimental setups to realize the
required SSH Hamiltonian with tunable parameters, and then discuss
how to measure the particle transport in the proposed cold atom
systems by numerically calculating the shift of the Wannier center
and the center of mass of an atomic cloud.

\subsection{Two experimental setups}

The first experimental setup we proposed is a 1D optical supperlattice trapping a noninteracting atomic gas of single-component fermions, as shown in Fig. 1(a). Such an optical lattice has been widely realized in experiments \cite{Porto2006,Bloch2007,Bloch2013,Bloch2014}. It is generated by superimposing two lattice potentials with short and long wavelengths differing by a factor of two, with the optical potential given by
\begin{eqnarray}
V(x)=V_{1}\sin^{2}(k_1x+\varphi)+V_{2}\sin^{2}(2k_1x).
\end{eqnarray}
Here $k_1$ is the wave vector of the short wavelength trapping lasers (the lattice spacing $a=2\pi/k_1$), $\varphi$ and $V_{1,2}$ are respectively the relative phase and the strengths of the two standing waves. By varying the laser intensity and the phase, one can fully control the lattice system with ease \cite{Porto2006,Bloch2007,Bloch2013,Bloch2014}, and then make the system well described by Hamiltonian (1) of SSH model in the tight-binging regime \cite{Bloch2013}. In the experiments \cite{Porto2006,Bloch2007,Bloch2013,Bloch2014}, the hopping configuration $J+(-1)^j\delta$ can be adjusted by varying potential strengths $V_{1,2}$ or swapping the relative phase $\varphi$, and the staggered potential $\Delta$ can be tuned by the phase. Therefore in this system, a straightforward way to realize the required hopping modulation with kink-type configuration in the time domain is by changing these tunable parameters of the optical superlattice with a well-designed sequence \cite{Bloch2013}.

Another experimental setup, which would be more convenient as we will see in the following, is loading an ultracold Fermi gas of two-component (internal states $|\sigma\rangle$ with $\sigma=\uparrow,\downarrow$) atoms in a state-dependent optical lattice \cite{Mandel}. It has been proposed to realize SSH Hamiltonian with a spatial domain wall in this system \cite{Ruostekoski1}, and such a state-dependent optical lattice have been experimentally created by superposing two linearly polarized laser beams with a relative polarized angle \cite{Mandel}. The separation and potential depth for different atomic components can be well controlled by the angle and the laser intensity, with a simple example of such a 1D lattice potential as shown in Fig. 1(b)
\begin{eqnarray}
V_{\sigma}(x)=V_{0}\sin^{2}(k_sx\pm \pi/4).
\end{eqnarray}
Here $V_0$ is the lattice potential depth, $k_s$ is the wavelength of the laser beams (the lattice spacing $a=2\pi/k_s$), and $\pm\pi/4$ are the polarization angles for atomic states $|\uparrow\rangle$ and $|\downarrow\rangle$, respectively.

For sufficiently deep lattices, the atoms in the system must alter their internal states in order to tunnel between two nearest-neighbor lattice sites. This can be achieved by the Raman-assisted tunneling method \cite{Jaksch,Gerbier,Bloch2014,Miyake,gauge_review}, as shown in Fig. 1(c). The energy offset between two atomic states arise from the external Zeeman field and then play the role of tunable parameter $\Delta$ in this system. Two pairs of Raman beams with Rabi frequencies $\Omega_{1,2}$ and $g_{1,2}$ are used to induce two large-detuned Raman transitions with detuning $\Delta_{d1}$ and $\Delta_{d2}$, respectively. One can use the former pair of Raman beams to realize the uniform nearest-neighbor hopping
\begin{eqnarray}
J=A_0\int w_{\uparrow}^{\ast}(x-x_{n}) e^{ik_xx} w_{\downarrow}(x-x_{n+1})dx,
\end{eqnarray}
where $A_0=|\Omega_{1}\Omega _{2}^{\ast}|/\Delta_{d1}$ is the effective Raman strength constant, $k_x$ is the momentum difference along the $x$ axis between the two beams, and $w_{\sigma}(x)$ are the Wannier functions of the lowest Bloch band for atomic state $|\sigma\rangle$.

To realize the time-varying hopping modulation term, one can use another pair of laser beams with a resulting Raman field, as shown in Fig. 1(b),
\begin{eqnarray}
R_{1}=g_{1}g _{2}^{\ast}/\Delta_{d2}=A_1(t)\sin(2k_sx+\pi/2),
\end{eqnarray}
where $A_1$ is a time-dependent constant controlled by the laser intensities or detuning $\Delta_{d2}$. In this way, the hopping modulation $(-1)^n\delta(t)$ is given by
\begin{eqnarray}
A_1(t)\int w_{\uparrow}^{\ast}(x-x_{n})\sin(2k_sx+\pi/2)w_{\downarrow}(x-x_{n+1})dx. \nonumber
\end{eqnarray}
Here the staggered hopping modulation is a consequence of the relative spatial configuration of the lattice and the Raman field: the period of $R_1(x)$ is double of the lattice period and $R_1(x)$ is antisymmetric corresponding to the center of each lattice site, as shown in Fig. 1(b). Thus, to realize the proposed particle transport scheme in this system, we can adjust the Zeeman field and the Raman field $R_0$ to tune the parameter $\Delta$ and $J$ in SSH Hamiltonian, and then independently vary the intensity of another Raman field $A_1(t)$ in time with a kink-type form. The case of time-varying $\Delta(t)$ can also be achieved in a similar way. The time modulation of the Raman coupling in ultracold atoms has been demonstrated in recent experiments \cite{Spielman2014}.

Considering $^{40}\text{K}$ atoms and typical lattice spacing $a=532$ nm, one has the recoil energy $E_R/\hbar\approx30$ kHz. For the optical superlattice system with intermediately deep lattice, a typical uniform hopping strength is $J\sim0.1E_R$, and the other parameters $\delta$ and $\Delta$ can be tunable in a wide regime \cite{Bloch2013}. For the state-dependent optical lattice system, the uniform hopping strength given by Eq. (13) is proportional to the effective Raman intensity $A_0$, which is typically in order of megahertz, and the overlap integral of Wannier functions between neighbor lattice sites can be about $10^{-2}$ \cite{Pachos}. Thus the Raman-induced uniform hopping strength in this system is $J\sim0.4E_R$, and the nature (next nearest-neighbor) hopping $t_N$ within sublattices can be effectively suppressed by sufficiently deep lattice $V_0$. For example, the numerical calculation shows that $t_N\lesssim10^{-3}E_R$ for $V_0\approx22E_R$ \cite{Pachos}. We can consider typical parameters $\delta_0$ and $\Delta$ (the minimum bulk gap in the dynamical progression is $2\Delta$) in the order of $0.1J$. In this case, the adiabatic approximation works well for the ramp time $t_f\gg\hbar/0.1J\sim1$ ms. Thus one can choose the ramp frequency $\xi=0.01J/\hbar$ and $t_0=50$ ms, which is well shorter than the typical coherence time in cold atom experiments. The non-adiabatic Landau-Zener transition from the ground band to the excited band for the chosen parameters is then given by $P_{\text{LZ}}\approx e^{-\pi(0.1J)^2/4\hbar\delta_0\xi}\approx0$. In addition, the finite temperature effects do not interfere with the particle transport progression for temperatures smaller than the energy gap \cite{TP1}. This requires the temperature of the order of $0.08E_R/k_B\sim 20$ nk ($k_B$ is the Boltzmann constant), which has been achieved in current experiments with, e.g., $^{40}\text{K}$ atoms. So we can conclude that the required Hamiltonian with tunable parameters and the adiabatic condition are able to be realized under realistic circumstances.

\subsection{Experimental measurement methods}

It has been shown  that the particle transport can be connected
with the Wannier center based on the modern theory of charge
polarization \cite{Vanderbilt,Xiao2010}. Especially, the shift of
the Wannier center in each unit cell is
\begin{eqnarray}
X_d= x_c(t=t_f)-x_c(t=0),
\end{eqnarray}
where  $x_c \equiv\langle w_n|\hat{x}|w_n\rangle$ is the Wannier
center, with $| w_n(x)\rangle=\frac{1}{2\pi}\int_{-\pi}^{\pi}
e^{-ik(n-x)}|u_-(k)\rangle$ being the Wannier function of the
ground band in the $n$-th unit cell. The shift of the Wannier
center encodes the adiabatic particle transport (the variation of
polarization) as \cite{TP1,TP4,TP5}
\begin{eqnarray}
X_d/a=Q.
\end{eqnarray}
In Fig. 3,  we have numerically calculated the variation of the
Wannier center in each lattice site in the proposed system. For
the symmetric case with $\Delta=0^+$ (the solid blue line) and
$\Delta(t)=\delta_0\sin(\pi t/2t_0)$ (the dashed red line) shown
in Fig. 3(a), after adiabatically tuning the hopping modulation
over the kink form, the Wannier center in each lattice site shift
downwards one site, that is, exactly one-half of the unit cell.
According to Eqs. (8) and (16), the adiabatic particle transport
is $Q=-1/2$ (the sign depends on the shift direction), as expected
for the half-charge in the original SSH model. For the
symmetry-breaking case with $\Delta=4\delta_0=0.1J$ as shown in
Fig. 3(b), the shift of the Wannier center is nearly half a
lattice site, which is consistent with the expected  $Q=-1/4$ in
this case.

\begin{figure}[tbp]
\includegraphics[width=8cm]{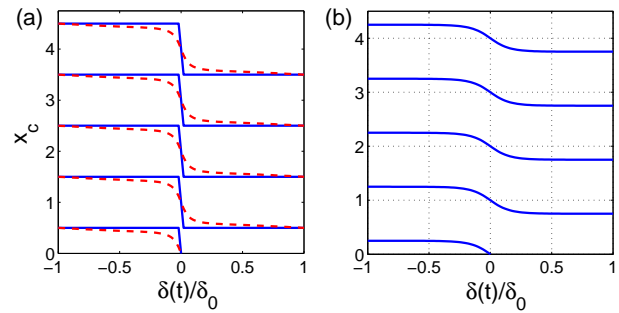}
\caption{(Color online) The shift of the Wannier center in each lattice as a response to the adiabatically tuning hopping modulation. (a) The symmetric case with $\Delta=0^+$ (solid blue line) and $\Delta(t)=\delta_0\sin(\pi t/2t_0)$ (dashed red line) with $\delta_0=0.1J$. After tuning the hopping modulation over the kink form, the Wannier center for each lattice site shift downwards one site, i.e. one-half of the unit cell, corresponding to $Q=-1/2$ as expected for the half-charge in the original SSH model. (b) The symmetry-breaking case with $\Delta=4\delta_0=0.1J$, the shift of the Wannier center is nearly half a lattice site, corresponding to $Q=-1/4$ expected in this case. Other parameters in (a) and (b) are $J=1$ and $t_0=5/\xi$.
}
\end{figure}

The shift of  Wannier center shown in Fig. 3 implies the
appearance of atomic current (the transport of particles) in each
unit cell, which flow through the whole lattice system. Under
the adiabatic condition, the atomic current will take the
solitonic form in the ramping progression, which is similar to the
example shown in Fig. 2(e). In principle, the transport dynamics
can be detected by using the single-atom {\sl in situ} imaging
technology in optical lattices \cite{insitu}. Thus, the variation
of atomic density distribution in a unit cell associated with
induced current can be experimentally extracted out in this way.
In our proposed systems, it would be more convenient to detect the
global current through the whole lattice by measuring the time
evolution of the atom fractions in the even and odd sublattices,
instead of using {\sl in situ} detection in a single unit cell.
For the double-well superlattice system in Fig. 1(a), the atomic
current associated with the atom fractions of the even/odd sublattices has been measured in
the experiment by transferring the atoms to higher-lying Bloch
bands and applying a subsequent band mapping technique
\cite{Bloch2014}. For the state-dependent optical lattice system
in Fig. 1(b), the even and odd sublattices trap $|\uparrow\rangle$
and $|\downarrow\rangle$ atoms, respectively. Therefore in this
system, one can simply measure the evolution of atom fractions via
optical imaging the up-component (down-component) atoms, such as
using state-resolved time-of-flight measurements \cite{tof}. Therefore, one can
obtain the particle transport $Q$ by its integration over the time
domain, as given by Eq. (10), which corresponding to the mimicked
FPN in this system.

\begin{figure}[tbp]
\includegraphics[width=8.6cm]{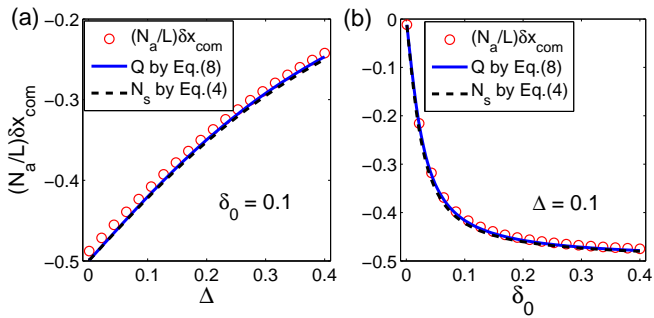}
\caption{(Color online) The shift of the center-of-mass of an
atomic cloud as a function of (a) the parameter $\Delta$ with
fixed $\delta_0=0.1$; and (b) the parameter $\delta_0$ with fixed
$\Delta=0.1$, respectively, for a finite lattice system with
$L=400$ sites and at half filling $N_a/L=1/2$. The solid (blue)
line and the dashed (black) line denotes the corresponding
particle transport $Q$ obtained by Eq. (8) and fractional particle
number $\mathcal{N}_s$ obtained by Eq. (4) in the text,
respectively. Other parameters in (a) and (b) are $J = 1$ and $t_0
= 5/\xi$.}
\end{figure}

In addition, the particle transport  can be directly measured from
the shift of the center of mass of an atomic cloud in a
finite lattice system \cite{TP1,TP2,TP5}. In our system, the center of mass
of an atomic cloud in the lattice with $L$ sites
$x_{\text{com}}(t)$ is given by
\begin{eqnarray}
x_{\text{com}}(t)=\frac{1}{N_a}\sum_{n=1}^{L}\sum_{\varepsilon_{oc}}|\psi_{oc}(n,t)|^2n,
\end{eqnarray}
where $N_a=L/2$ is the atomic number  at half filling,
$\varepsilon_{oc}$ denotes the occupied state of the fermionic
atoms, and $\psi_{oc}$ is the corresponding wave function. Under
the adiabatic evolution with $\delta(0)\rightarrow\delta(t_f)$
described by Eq. (6), the center of mass of the system shift from the
position $x_{\text{com}}(0)$ to $x_{\text{com}}(t_f)$. It can be
proved that the shift of the center of mass $\delta
x_{\text{com}}=x_{\text{com}}(t_f)-x_{\text{com}}(0)$ is
proportional to the particle transport in the infinite $L$ limit
\cite{TP1,TP5}
\begin{eqnarray}
\frac{N_a}{L}\delta x_{\text{com}}=Q.
\end{eqnarray}
If $L$ is large enough, such that the  bulk properties of the
system are almost not affected by the edges, $N_a\delta
x_{\text{com}}/L$  in the above equation will be approximated to be
the ideal particle transport $Q$ in infinite system. In Fig. 4, we
have calculated the shift of the center of mass of an atomic cloud for a
lattice system with $L=400$ sites, $N_a\delta
x_{\text{com}}/L$ (the red circles) as a function of the
parameters $\Delta$ [Fig.4 (a)] and $\delta_0$ [Fig.4 (b)],
respectively. As shown in Fig. 4(a) and 4(b), the calculated shift
of the center of mass is well described by the particle transport $Q$ (the
solid blue line) obtained by Eq. (8) with small deviations. These
deviations are due to the finite size effects and become smaller
and smaller with the increase of the lattice size in our simulations, which will be invisible in practical
experiments. For comparison, we also plot the mimicked FPN
$\mathcal{N}_s$ given by Eq. (4) in this system in Fig. 4 (the
dashed black line). From Fig. 4, one can see that the corresponding FPN
is nearly equal to the particle transport within the parameter
regimes. In current experiments, the center-of-mass position of an atomic cloud
can be directly and precisely measured, either by using {\sl in
situ} measurement of the atomic density distribution in the optical lattices \cite{insitu} or
deduced from the time-of-flight imaging \cite{tof}.

Finally, we note that a shallow-enough harmonic trap in practical experiments will not affect the particle transport \cite{TP1,TP4} and the main results of this paper remain intact. In order to take the effect of the harmonic trap into account, we can add a term
$H_t=V_t\sum_n(n-L/2)^2\hat{c}^{\dag}_n\hat{c}_n$ into Hamiltonian (1), where $V_t$ is the trap strength and $L$ is the
lattice size. Within a local density approximation, the lower band will still be filled at the center of the trap and thus the shift of the Wannier center
in these lattice sites is expected to be nearly the same as those shown in Fig. 3; while the band is only partially filled near the edge with the local trap energy $V_t(n-L/2)^2\gtrsim E_g$, such that the pumping argument does not apply to this region \cite{TP1}. Therefore in practical experiments, one may emphasize on the shift of Wannier center in the central region or turn the trap strength to a small value $V_t\sim 4E_g/L^2$. For the shift of the center of mass of an atomic cloud, our numerical simulations demonstrate that the results shown in Fig. 4 preserve with a deviation less than $2\%$ for $V_t\approx0.6\times10^{-5}J$, while $V_t\approx10^{-4}J$ is enough if the lattice size reduces to $L=100$, which are consistent with the estimates in the local-density analysis.

\section{conclusions}

In summary, we have proposed a scheme to mimic and measure the
FPN in the generalized SSH model with cold fermions in 1D optical
lattices. It has been shown that FPN in this model can be simulated in
the momentum-time parameter space in terms of Berry curvature
without a spatial domain wall. In this simulation, a hopping
modulation is adiabatically tuned to form a kink-type
configuration and the induced current plays the role of soliton in
the time domain, so FPN is expressed by the particle
transport. We have also proposed two experimental setups of
optical lattices to realize the required Hamiltonian with tunable
and time-varying parameters, and considered the energy scales and
the adiabatic condition under practical circumstances. Finally we
have discussed how to directly measure the particle transport in
the proposed systems by numerically calculating the shift of the
Wannier center and the center of mass of an atomic cloud. Considering that
all the ingredients to implement our scheme in optical lattices
have been achieved in the recent experiments, it is anticipated
that the presented proposal will be tested in an experiment in the
near future. The direct measurement of such a mimicked FPN in cold
atom experiments will be an important step toward exploring
fractionalization and topological states in cold atom systems. Extensions of this work can enable to simulate and measure FPN emerging in two- and three-dimensional Dirac Hamiltonian with topologically nontrivial (vortex and monopole) background fields \cite{JR,2D-FPN}, which has been theoretically studied but remains elusive in nature. It will be also interesting to simulate a variety of topological states \cite{Qi2008a} and study their properties in the parameter space using cold atoms.

\section{Acknowledgements}

This work was supported by the RGC of Hong Kong (Grants No.
HKU7045/13P and HKU173051/14P), the SKPBR of China (Grants No.
2011CB922104 and 2013CB921804), the NSFC (Grants No. 11125417 and
11474153), and the PCSIRT (Grant No. IRT1243).

\end{document}